\documentclass[journal]{IEEEtran}

\usepackage{graphicx}
\usepackage{amssymb}

\begin{document}
\title{Surface wave control for large arrays of microwave kinetic inductance detectors}
\author{Stephen J. C. Yates, Andrey M. Baryshev, Ozan Yurduseven, Juan Bueno, Kristina K. Davis, Lorenza Ferrari, Willem Jellema, Nuria Llombart, Vignesh Murugesan, David J. Thoen, Jochem J. A. Baselmans.
\thanks{This work was in part supported by ERC starting grant ERC-2009-StG Grant 240602 TFPA. The contribution of J.J.A. Baselmans is also supported by the ERC consolidator grant COG 648135 MOSAIC. This work was part of a collaborative project, SPACEKIDs, funded via grant 313320 provided by the European Commission under Theme SPA.2012.2.2-01 of Framework Programme 7.}
\thanks{Stephen Yates, Andrey Baryshev, Lorenza Ferrari, and Willem Jellema are with SRON, Landleven 12, 9747 AD Groningen, The Netherlands.}
\thanks{Jochem Baselmans, Juan Bueno, and Vignesh Murugesan are with SRON, Sorbonnelaan 2, 3584 CA Utrecht, The Netherlands.}
\thanks{Ozan Yurduseven, Nuria Llombart, Jochem Baselmans, and David Thoen are in the Terahertz Sensing Group, Faculty of Electrical Engineering, Mathematics and Computer Science, Delft University of Technology, Mekelweg 4, 2628 CD Delft, The Netherlands.}
\thanks{Andrey Baryshev and Willem Jellema are with the Kapteyn Astronomical Institute, University of Groningen, P.O. Box 800, 9700 AV Groningen, The Netherlands.}
\thanks{Kristina Davis is with the Arizona State University, 781 Terrace Rd., Tempe, AZ, U.S.A}}
\maketitle 
\begin{abstract}
Large ultra-sensitive detector arrays are needed for present and future observatories for far infra-red, submillimeter wave (THz), and millimeter wave astronomy. With increasing array size, it is increasingly important to control stray radiation inside the detector chips themselves, the surface wave. We demonstrate this effect with focal plane arrays of 880 lens-antenna coupled Microwave Kinetic Inductance Detectors (MKIDs). Presented here are near field measurements of the MKID optical response versus the position on the array of a reimaged optical source. We demonstrate that the optical response of a detector in these arrays saturates off-pixel at the $\sim-30$~dB level compared to the peak pixel response. The result is that the power detected from a point source at the pixel position is almost identical to the stray response integrated over the chip area. With such a contribution, it would be impossible to measure extended sources, while the point source sensitivity is degraded due to an increase of the stray loading. However, we show that by incorporating an on-chip stray light absorber, the surface wave contribution is reduced by a factor $>$10. With the on-chip stray light absorber the point source response is close to simulations down to the $\sim-35$~dB level, the simulation based on an ideal Gaussian illumination of the optics. In addition, as a crosscheck we show that the extended source response of a single pixel in the array with the absorbing grid is in agreement with the integral of the point source measurements.
\end{abstract}
\begin{IEEEkeywords}
 microwave kinetic inductance detector, KID, antenna, low temperature detector, surface wave, twinslot, submillimeter wave, terahertz.
\end{IEEEkeywords}
\IEEEpeerreviewmaketitle
\section{Introduction}
\begin{figure}
\centering
\includegraphics*[width=8.5cm]{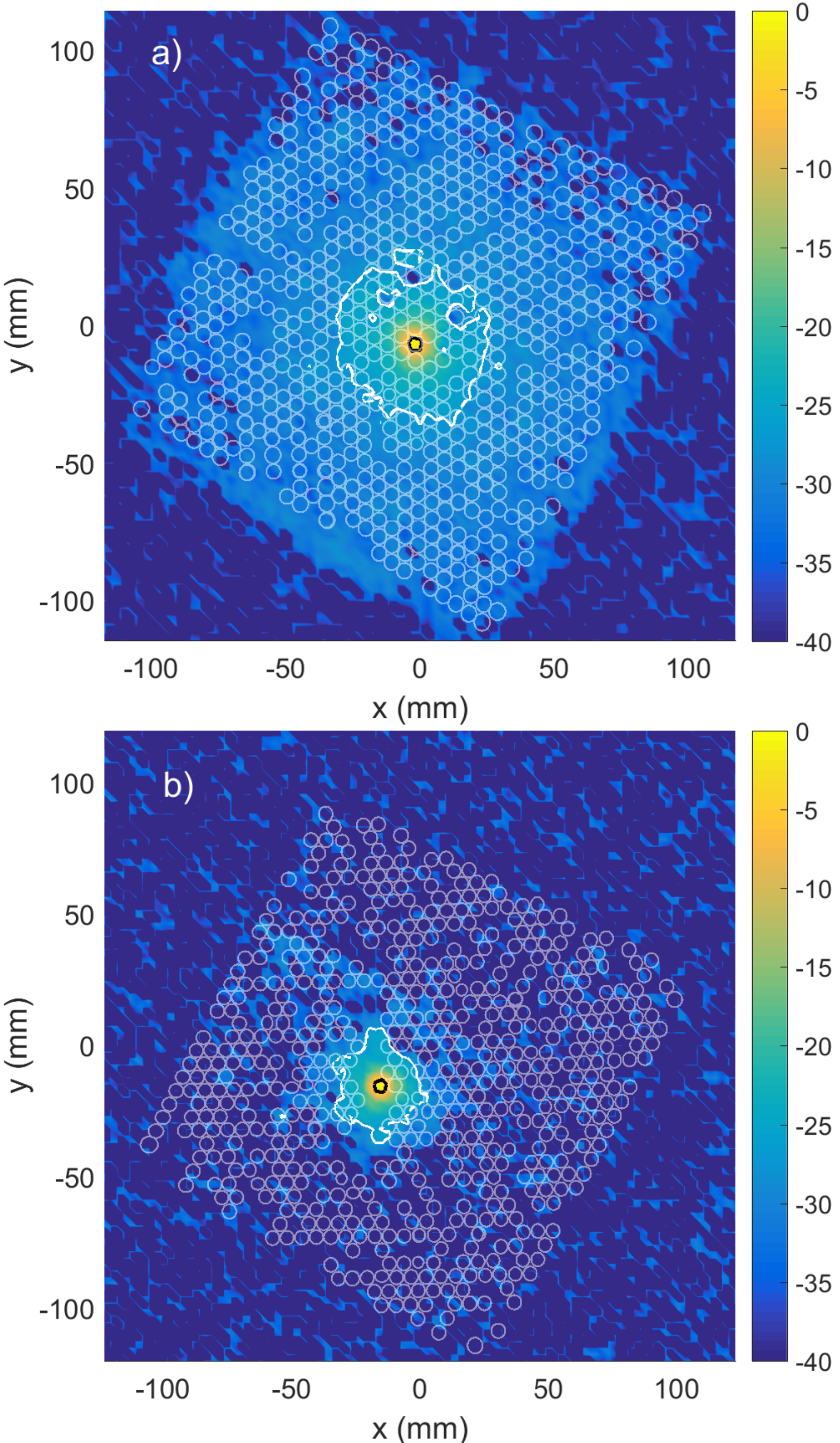}
\caption{Position dependent response in dB of one pixel to a point source placed in a reimaged focal plane with a magnification of 3. The $-3$~dB and $-27$~dB contours are shown. The circles show the fitted 3~dB beams of all found pixels, shown to show the extent of the array. Two arrays are shown: a) without on chip stray light absorbing mesh; b) with absorbing mesh. Note the large area response at the $\sim-30$~dB level without the mesh disappears on the array with the on chip absorbing mesh.}
\label{fig:hot_beams}
\end{figure}

\IEEEPARstart{P}{resent} and future observatories for far infra-red (FIR,$\sim$1--10~THz), submillimeter wave (0.3--1~THz) and millimeter wave (50--300~GHz) astronomy need increasingly large arrays of ultra-sensitive power (``direct'') detectors~\cite{baselmans:AA17}. This requires a CCD-like approach in which large scale monolithic detector chips are combined with a multiplexed readout. Current imaging arrays for the far infra-red and the submillimeter regime are based upon transition edge sensors (TES's) \cite{Bock1995} or Microwave Kinetic Inductance Detectors (MKIDs) \cite{day03}. In both cases the detector arrays are based upon large, monolithic chips, where radiation coupling is achieved using planar absorbers, lenses or horns. With ever increasing array size it becomes critically important to control stray radiation inside these detector chips. Even in the best cases the radiation absorption in a single pixel is not perfect: part of the radiation can be reflected and re-scattered into the dielectric of the detector chip. This confined radiation is commonly referred to as a surface wave. Typical chip materials such as Si have a high refractive index, increasing the probability of total internal reflection. To illustrate the effect we show in Fig.~\ref{fig:hot_beams}(a) the spatial response of a central pixel of an 880 pixel array of lens-antenna coupled MKIDs as a function of the position of a small calibration source in the image plane of the chip, the system beam pattern. We observe a localized peak response, the main beam,  at the pixel position. However, we also observe a low-level of response over the entire chip area, which we will refer to as the ‘pedestal’ response in the remainder of the text. The pedestal response consists of power coupled to the chip at a position spatially far away from the measured pixel: it is detected at the pixel under test due to scattering of radiation inside the detector chip. Normalizing the system beam pattern to its maximum response, the pedestal response is seen at a level of $\sim-30$~dB. In this particular case the total integrated stray power in the pedestal at $-30$~dB is similar to the power in the main beam. This will render imaging of extended sources impossible and results in excess power loading when using this array for ground based astronomy. In this paper we study this problem in detail by comparing two large imaging arrays, which are based upon lens-antenna coupled MKIDs. Both arrays are identical with the exception of an absorbing mesh layer designed to absorb re-scattered radiation propagating through the detector chip, with the resultant pedestal-suppressed spatial response shown in Fig.~\ref{fig:hot_beams}(b). We discuss in detail the design, fabrication and testing of these two systems and demonstrate that the aforementioned problem can be reduced very significantly by using a stray-light absorbing layer. 

\begin{figure*}[hbt]
\centering
\includegraphics[width=1\textwidth]{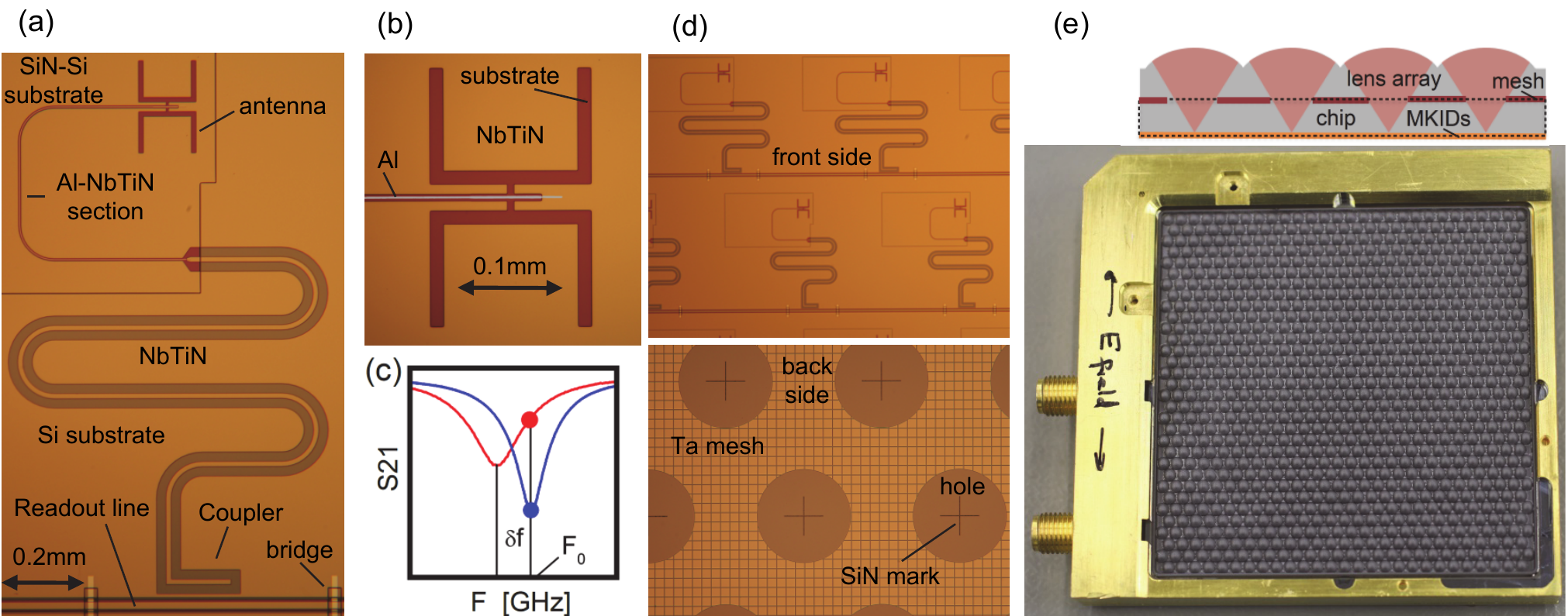}
\caption{(a) Optical micrograph of a single pixel of the array, artificial coloring is used to highlight the different metals.  (b) Zoom in to the antenna structure of panel (a). (c) The transmission of the readout line around a single MKID, measured at 2 different values of the power absorbed by the device, showing the response mechanism of the MKID (d) Optical micrograph of the array, with the front side (top) and backside (bottom) showing the detectors on the front side and the Ta absorbing mesh on the backside, implemented on only one of the two arrays discussed in the text. (e) Assembled detector holder with lens array and SMA connector for contacting the readout circuitry. The top panel shows schematically the assembled cross section.}
\label{fig:KID_photo}
\end{figure*} 

\section{Array design}
\begin{figure*}
\centering
\includegraphics*[width=1\textwidth]{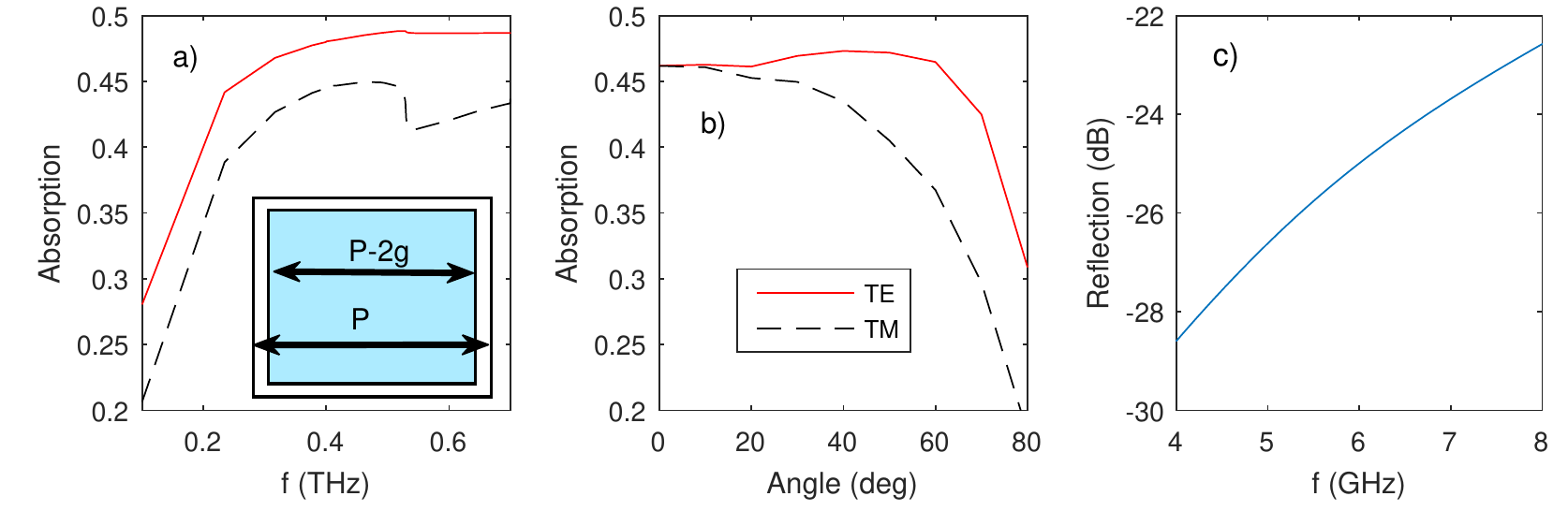}
\caption{Calculated mesh performance: a) the frequency dependence of the absorption of the transverse electric (TE) and transverse magnetic (TM) modes at 40 degrees incidence to the mesh. Insert, the mesh unit cell design, blue mesh, white the substrate. Values used of $P=120~\mu$m and $g=3.5~\mu$m; b) the angular dependence at 350~GHz. c) Reflection of the mesh in the MKID readout band.}
\label{fig:mesh_cal}
\end{figure*} 

\begin{figure}
\centering
\includegraphics*[width=8.5cm]{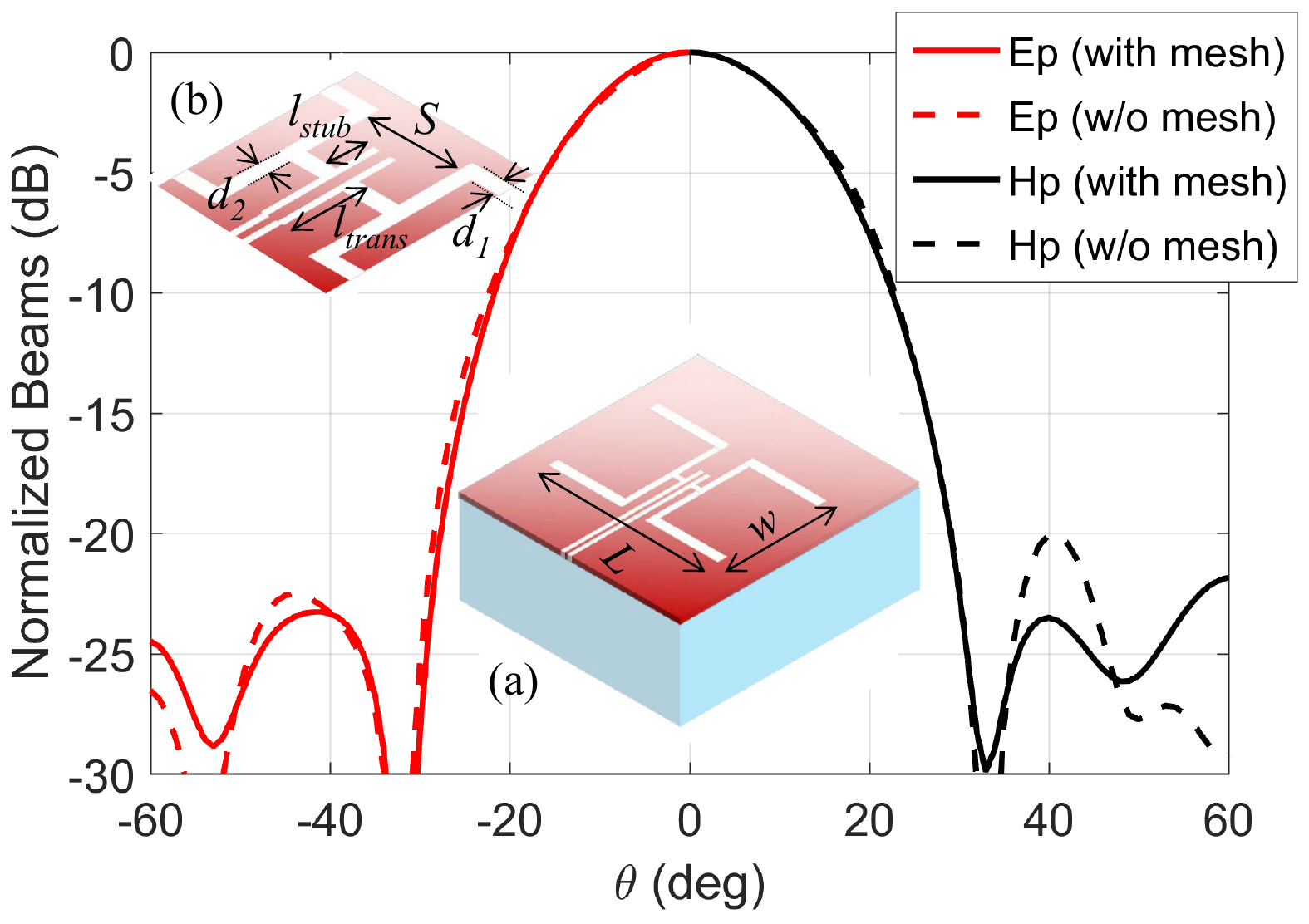}
\caption{Calculated far field detector beam pattern, with and without absorbing mesh absorber. Ep is the E plane, perpendicular to the antenna slots; while Hp is the H-plane, parallel to the antenna slots. Inset (a) shows the antenna geometry and (b) and zoom on the antenna feed, transformer and stub, see text or \cite{Ozan2016} for details.}
\label{fig:CST_beam}
\end{figure}

The arrays discussed in this paper are based upon NbTiN-Aluminum lens-antenna coupled MKIDs, similar to the device discussed by Janssen et al. \cite{Janssen:APL13}. A micrograph of a single MKID of the array is given in Fig. \ref{fig:KID_photo}(a). The device is fabricated on a 350~$\mu$m thick Si $<100>$ FZ wafer with a resistivity $\mathrm{\rho>10~k\Omega cm}$. Additionally, both sides of the wafer are coated with a 0.5~$\mu$m layer of (low pressure chemical vapor deposition) LPCVD SiN. Each detector consists of a meandering coplanar waveguide (CPW) with an open end near the readout line and a shorted end at the location of the antenna. The MKID is made out of a 500~nm thick film of NbTiN, deposited using reactive magnetron sputtering in an argon-nitrogen plasma \cite{Boy2016,Thoen2016}. The device has a wide section with a central line width = 14~$\mu$m and a central line to ground plane gapwidth = 24 $\mu$m made from NbTiN. Here, the SiN layer is removed prior to the NbTiN deposition to reduce excess device noise due to two level systems associated with the amorphous SiN~\cite{Gao2008b}. The last $\sim$1.5~mm section of the MKID is narrow and here the central line of the CPW is made out of 55~nm sputter deposited aluminum (linewidth = 1.6 $\mu$m, gapwidth = 2.2~$\mu$m) to enhance the device response and optical efficiency. The SiN is present here to increase the device yield at the Al/NbTiN interface~\cite{Ferrari:inprep}. The MKID is read out by using a readout signal via the CPW through-line at a single frequency corresponding to the first distributed resonance occurring at a frequency $F_0=\frac{c}{4L\sqrt{\epsilon_{eff}}}$. Here {\it L} is the resonator length, {\it c} the speed of light and $\epsilon_{eff}$ the effective dielectric constant of the CPW. Radiation coupling to the devices is achieved by a twin-slot antenna~\cite{Jonas:IEEETMTT92,filipovic:IEEETMTT93}, coupled to the shorted end of the resonator as shown in Fig.~\ref{fig:KID_photo}(b). The antenna is optimized for radiation coupling in a 60~GHz band around 350~GHz~\cite{Ozan2016}. The geometry of the antenna, impedance matching stub and transformer~\cite{Ozan2016} are indicated by the insets Fig.~\ref{fig:CST_beam}, where the parameters shown are: L=240~$\mu$m; W=137~$\mu$m; d=12~$\mu$m; S=25~$\mu$m; $\mathrm{l_{stub}}$=26um; $\mathrm{l_{trans}}=40~\mu$m. Additionally, the CPW in the stub and transformer have a central line of width $2~\mu$m with a gap to the ground plane of $2.2~\mu$m; this transitions to a CPW of width $1.6~\mu$m with a gap of $2.2~\mu$m for the Al section of the MKID. The angular dimension of the lens is designed such that 82~\% of the power is captured by the lens. This efficiency estimates how much is the power launched into the surface wave from a single double slot antenna.  However, the MKID CPW line will itself directly weakly couple to the surface wave and therefore increase the contribution to the detector from the surface wave. Radiation coupled to the antenna is transferred to the narrow NbTiN-Al CPW line of the MKID and absorbed only in the aluminum central strip of the MKID: the gap frequency of NbTiN does not allow for radiation absorption below 1.1~THz whereas aluminum absorbs radiation for frequencies in excess of 90~GHz. The consequence is that there is no radiation loss in the device and thus there is a very high detector efficiency~\cite{Janssen:APL13}. The result of the radiation absorption is that the MKID resonant frequency shifts to lower frequencies and that the MKID resonance feature broadens (see Fig.~\ref{fig:KID_photo}(c)).

The two arrays we consider in this paper both consist of 880 pixels hexagonally packed with a pixel spacing of 2 mm, covering an area of $55.7\times56$~mm on a $62\times60.8$~mm chip. In Fig.~\ref{fig:KID_photo}(d) we show a combined micrograph of part of the array front side and back side. Across the array, the MKID length {\it L} is changed systematically from 6.6 to 3.5~mm, resulting in F$_0$ ranging from 4.2 to 7.8~GHz. Note that all devices are coupled to a single readout line; electrical contact to the chip is achieved by just two bond pads. The SiN layer is present below the central conductor of the readout line to allow electrical measurements of the readout-line integrity at room temperature during the fabrication process; without this layer the Si wafer will short-out the series resistance of the NbTiN line. The presence of the SiN below the readout line does not change its loss tangent: it is measured to be $\tan \delta \sim 5\times10^{-5}$ using a test resonator, allowing a good signal coupling to the low noise amplifier. We use aluminum bridges with lithographically-defined polyimide supports to balance the two grounds of the readout line. Additionally, we spatially encode the pixels such that neighboring pixels are separated sufficiently in readout frequency. Both techniques reduce MKID-MKID crosstalk~\cite{Yates:JLTP14}. Residual crosstalk is now limited by resonator overlap~\cite{Adane:JLTP16,Bisigello:SPIE16,baselmans:AA17}, limited in our case by the NbTiN film flatness~\cite{Thoen2016} and the MKID Q-factors under operation. Efficient radiation coupling to the MKID antennas is achieved by using a Si lens array of spherical lenses fabricated using laser ablation from a separate Si wafer. The lens array and chip are mounted together using a dedicated alignment and bonding technique where the lens array and chip are pressed together using a silicone-based press system before a semi-permanent bond is made using Locktite 406 glue. This method guarantees a glue gap below 5~$\mu$m over the entire chip area. Alignment is achieved by markers in the SiN layer on the detector chip backside that were etched in the first step of the device fabrication (see  Fig.~\ref{fig:KID_photo}(d)). The large area of the chip requires the lens array and the detector chip to be made from the same material to guarantee reliable bonding during thermal cycling of the detector assembly. The detector chip is mounted in a dedicated holder and wire bonding is used to contact the two bond pads to standard SMA co-ax connectors, the finished assembly is shown in Fig.~\ref{fig:KID_photo}(e).

\begin{figure}
\centering
\includegraphics*[width=8.5cm]{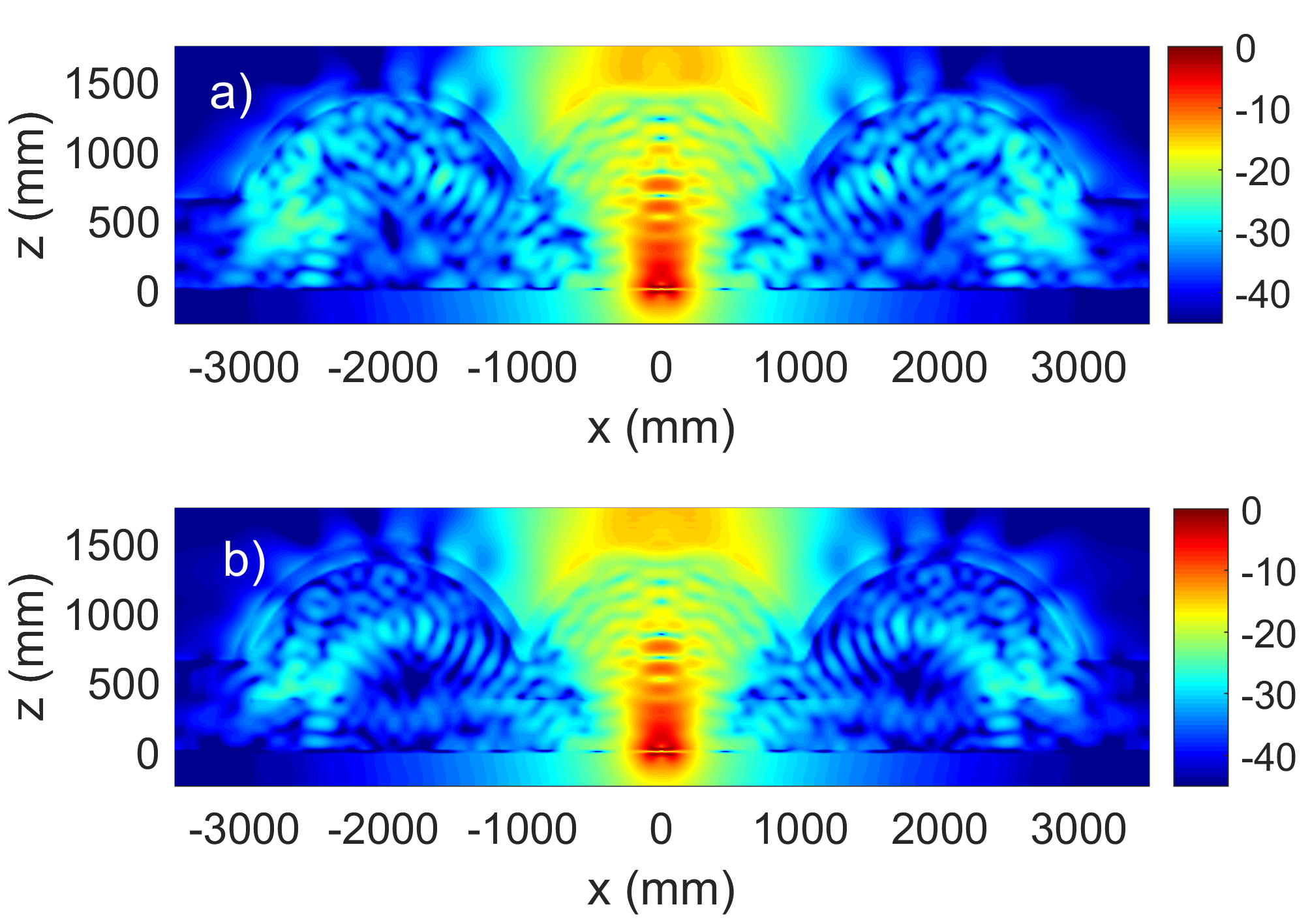}
\caption{Cross cuts of simulations~\cite{CST} of the electric field strength (in dB) for the lens-antenna system including nearest neighbors pixels, showing leakage of power into the surface wave. a) is without absorbing mesh, b) with. With the mesh the about half of the power in the surface wave is absorbed within an distance of a single lens.}
\label{fig:CST_cuts}
\end{figure}

The second array is equipped with a stray radiation absorbing layer, on the backside of the detector chip, fabricated from a 40~nm thick Ta layer deposited using DC magnetron sputtering at room temperature. Under these growth conditions Ta grows in its $\beta$-phase \cite{schrey1970}, characterized by a high resistivity and low critical temperature \cite{mohazzab2000}. For our film we measure a sheet resistance $\mathrm{R_s=61~\Omega/\square\;and\;T_{c}=0.65~K}$; it is noteworthy that this observed sheet resistance gives the maximum radiation absorption for a metal layer in between two Si substrates (i.e. the lens array and the chip). The gap frequency of the Ta layer is approximately 50~GHz, that is at the readout frequency the material is superconducting and at 350~GHz it is resistive with a resistivity very close to the normal state resistance. Using a parametric sweep over parameters P and g (Fig.~\ref{fig:mesh_cal}(a), inset), the mesh design is optimized for maximum radiation absorption upto large angles for both the TE and TM mode at 350~GHz and for maximum transmission at the MKID readout frequency of 4--8~GHz; the optimized curves are shown in Fig.~\ref{fig:mesh_cal} together with a zoom of the mesh structure. The transparency from 4--8~GHz is needed because the mesh is only 350~$\mu$m distance from the MKIDs and therefore close enough to couple to the device. Without this the MKID will be sensitive to power absorbed in the stray light absorbing layer and would additionally have an enhanced coupling to the readout line. To efficiently couple the radiation from the lenses to the antenna, a 1.1~mm diameter hole is etched in the mesh; this is shown in Fig.~\ref{fig:KID_photo}(d) and the inset of panel (e). The far field lens-antenna beam pattern is simulated~\cite{CST}, with and without mesh present and shown in Fig.~\ref{fig:CST_beam}. This shows the mesh has a small perturbation on the beam pattern, reducing the calculated lens-antenna aperture efficiency from 0.75 to 0.74. To simulate the surface wave the case of a small 7 pixel array is taken, the electric field strength~\cite{CST} is shown in Fig.~\ref{fig:CST_cuts}. From this calculation, we determine that the mesh absorber reduces the amount of power in the surface wave by absorbing about half the power within a distance of a single lens (2~mm) from the antenna.

As a last fabrication step we thermally evaporate a layer of Ti-Cu-Au (5, 500 and 100~nm) layer at the chip edge, exactly where it is pressed into the sample holder. The role of this layer is to strongly increase the thermal contact between the detector chip and the holder. With a thermalization layer there is no residual bath-T dependent thermal response measurable at 240~mK. 

\section{System beam pattern}
\begin{figure*}
\centering
\includegraphics*[width=1\textwidth]{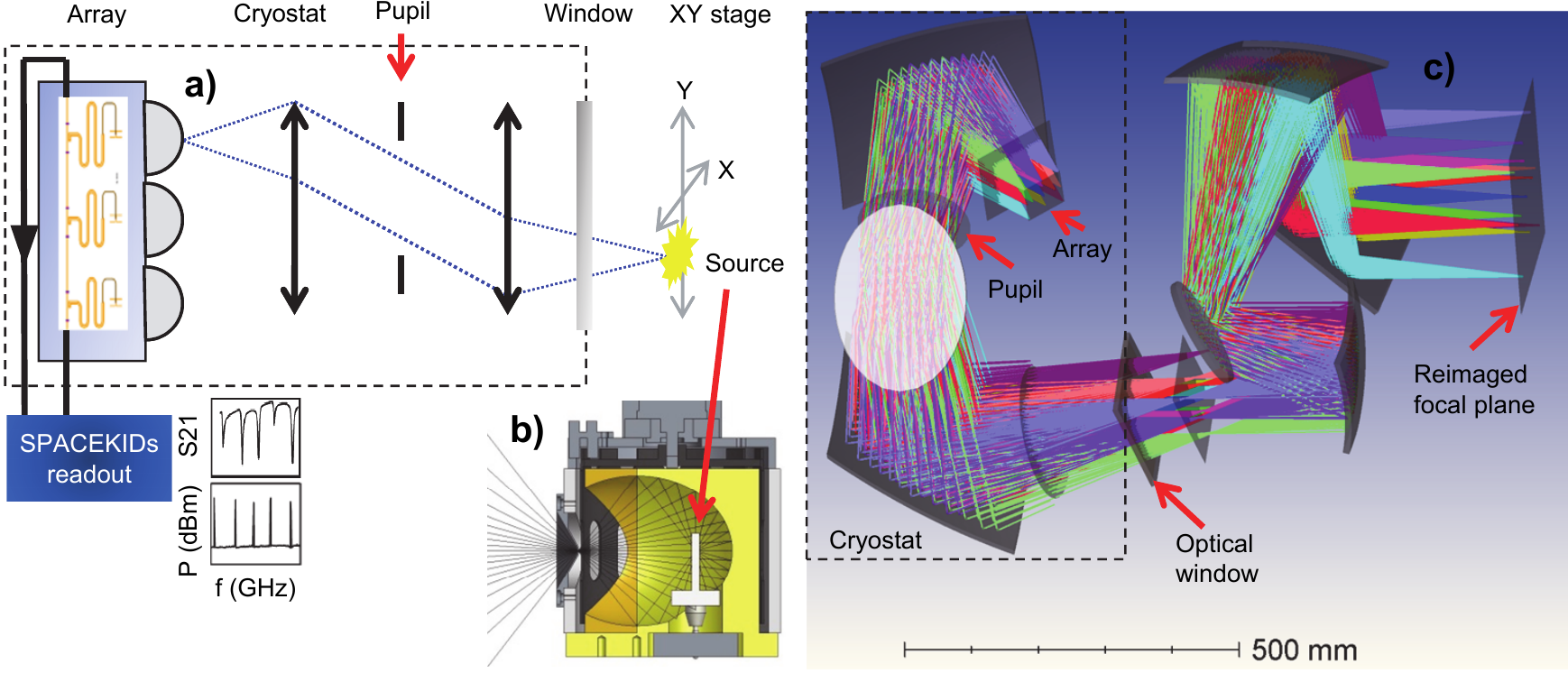}
\caption{a) A simplified schematic of an optical relay used in the measurement system, showing how a source is reimaged on the array to measure the position dependent response. The arrows give the mirror positions and the dotted lines indicate the optical beam. The full system has two such relays back to back, with 4 active mirrors and 3 fold mirrors. Indicated is the pupil, which limits the angular range and sets the spatial sampling of the system diffraction pattern. Inset lower left is the example MKID transmission and tones used to read them out using the ``SPACEKIDs'' readout~\cite{rantwijk:IEEE16}. b) cross-section of hot source assembly. c) Actual ray trace showing positions of the main apertures and all mirrors.}
\label{fig:schema}
\end{figure*} 

\begin{figure}
\centering
\includegraphics*[width=8.5cm]{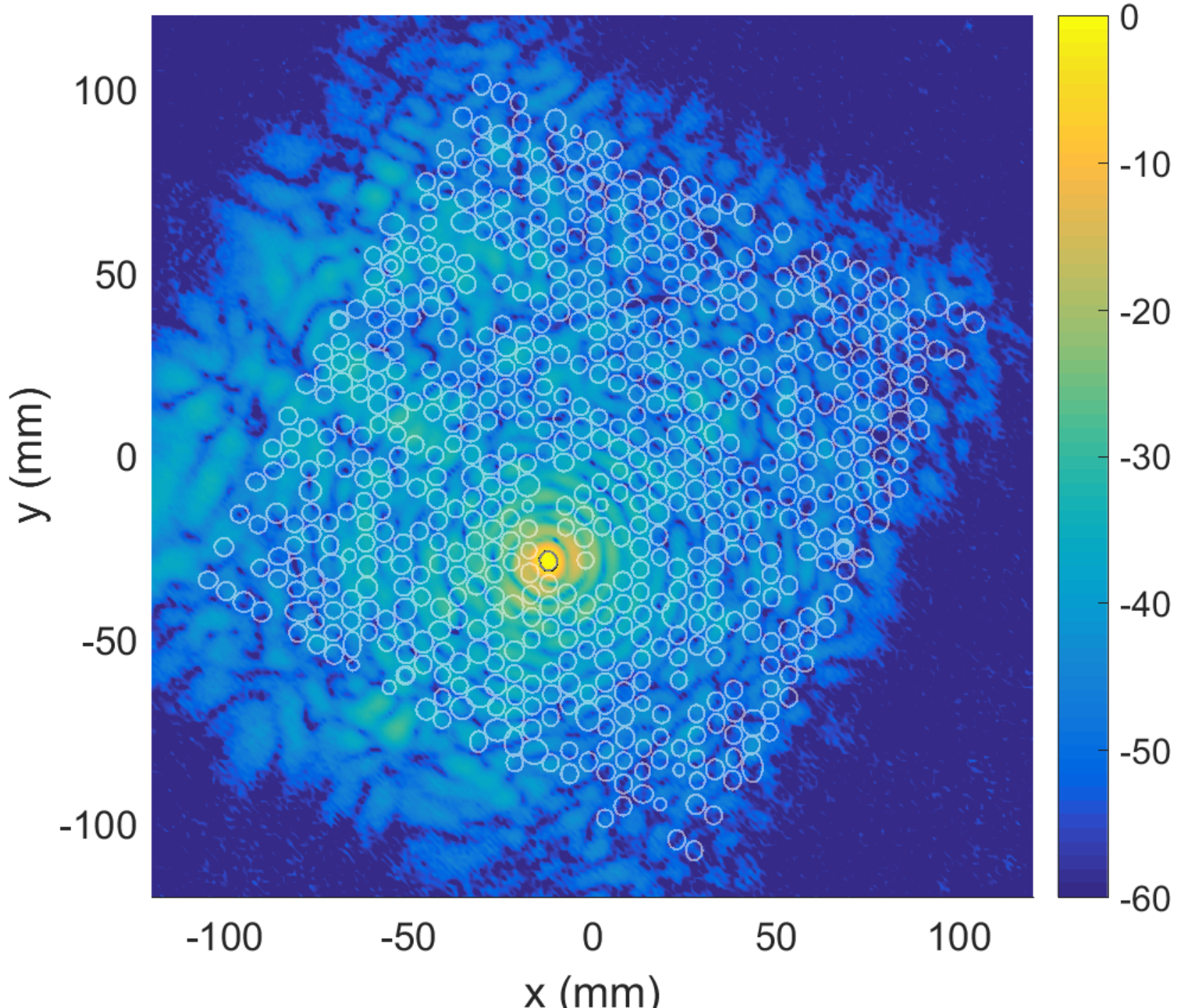}
\caption{Beam pattern for the array with an on-chip stray light absorber measured using the phase and amplitude method, scaled in dB. 3dB contours of all pixels are also shown so indicating the array size. The array is smaller than the field of view (FOV), the FOV edge is seen where the signal drops into the noise floor.}\label{fig:PA_beam}
\end{figure} 

\begin{figure*}[th]
\centering
\includegraphics*[width=17cm]{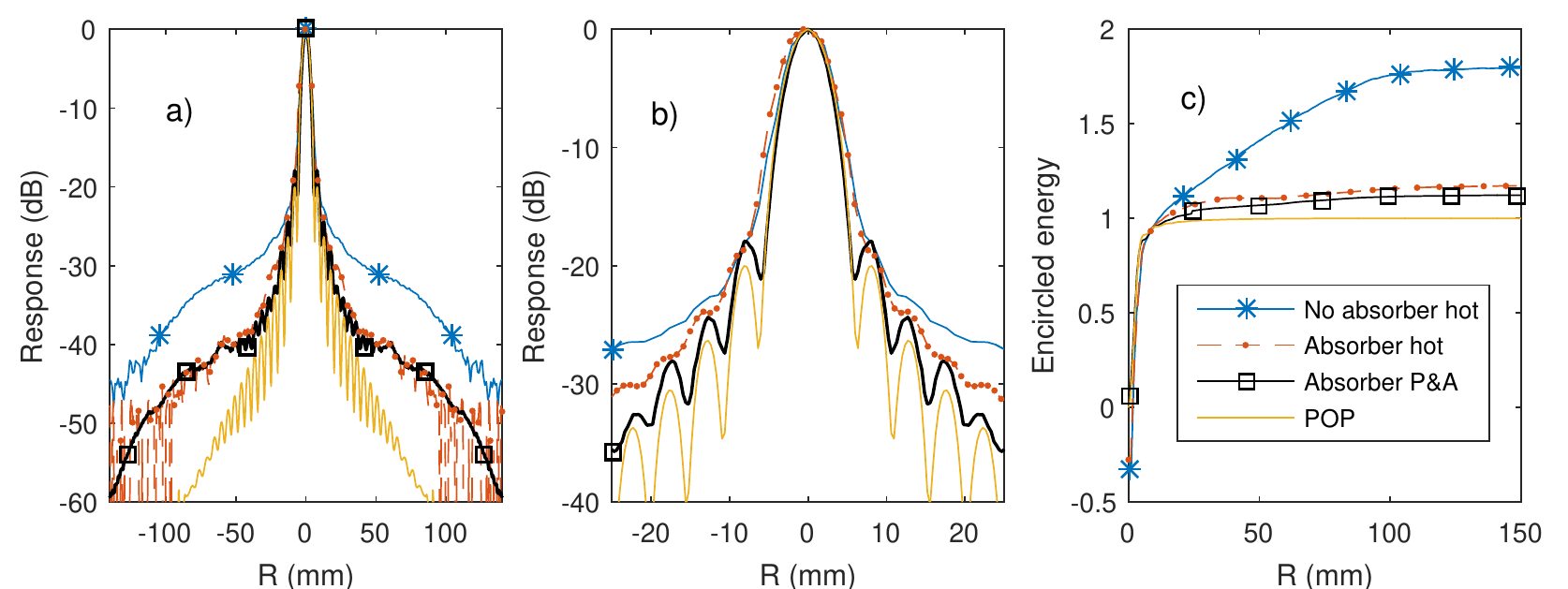}
\caption{a) Radial mean of beam pattern. Shown is the median of $\sim20$ pixels from the array center. b) Zoom on the radial mean. c) Encircled energy, the integral of the beam pattern to a given radius centered on each pixel. The encircled energy is the median value of $\sim$ 20 central pixels, normalized to the total power of the POP simulation and corrected for slightly different beam radii (see text for details).}\label{fig:radial_mean}
\end{figure*}

The arrays were measured individually in a submillimeter wave camera cryostat, with the arrays mounted on a thermal isolation suspension connected to the 240~mK stage of a three stage He$^{3}$/He$^{3}$/He$^{4}$ sorption cooler. The two additional cold stages of this three stage cooler are used to thermally buffer the co-axial readout lines and thermal-mechanical suspension holding the detector assembly. The camera optics create an image of the detector array at a warm focal plane outside the cryostat using a seven mirror system with a system magnification of 3. The optical design is based on two back-to-back optical relays, each consisting two off-axis parabolic mirrors forming a Gaussian beam telescope~\cite{Goldsmith}, shown in Fig.~\ref{fig:schema}(a)~\&~(b). One of the optical relays is placed at 4~K and the other outside the cryostat. The optical design is based on aberration compensation~\cite{murphy:IJRMW87}, canceling the aberrations and cross-polarization of the optics near the optical axis. To improve performance over the entire (large) field of view the mirror shape and angles are optimized, giving a low distortion, diffraction limited performance with a Strehl ratio of greater than 0.97 across the entire field view at 350~GHz and even at 850~GHz. Three fold mirrors are used to minimize the total size of the optics system and give a horizontal beam with a usable warm reimaged focal plane. This rotates the focal plane, which is not corrected for in the presented data. An angular limiting aperture ``the pupil'' limits the beam to a focal length to beam diameter (f-number or f\#) of f\#=2 and is placed between the 4~K active mirrors where all the different pixel beams overlap. The designed beam truncation at the pupil is $\sim-3$~dB. The arrays sample the focal plane with 2~mm pixels at 350~GHz, a spatial sampling of $\sim$ 1.2f\#$\lambda$. The design was simulated using the Zemax physical optics (POP) tool,~\cite{Zemax} shown in Fig.~\ref{fig:radial_mean}. Using these simulations, all mirror sizes were designed with low spillover $<-20$~dB, to couple the designed beam efficiently to the warm focal plane. 

The total power entering the cryostat window from room temperature is 70~dB larger than the power admitted to the detector chip. Part of this power is out-of-band radiation, mainly infra-red (IR), and part is due to the much larger throughput into the cryostat window compared to the cold optics. We achieve this 70 dB rejection by using the concept of a box-in-a-box, where each later, colder and lower power part of the optics is enclosed in a separate baffled light-tight box with a filtered optical window. The large field of view requires filters that are very large, with sizes up to $\sim\varnothing$20~cm. The large angular optical throughput and field of view give a very large IR thermal load, which combined with the poor thermal conductance of these filters results in significant filter heating ~\cite{Tucker:SPIE006}. To limit this we use a set of reflective and scattering IR filters at 300~K, 50~K and 4~K. The transmission band is further reduced with low-pass filters at 50~K and 4~K. Even then the 50~K stage low-pass filter is expected~\cite{Tucker:SPIE006} to be $>$150~K while the 4~K low-pass filter was measured to be 32~K in its center. Additional IR blocking, low-pass and band-pass filters at 4~K, 800~mK and 240~mK ensure efficient absorption or reflection of this radiation. The bandpass also defines the measurement band as the antenna itself has a wider bandwidth than necessary here.

The arrays are read out using an in-house developed multiplexed readout system~\cite{rantwijk:IEEE16}, which allows 2~GHz of readout bandwidth around a central local oscillator (LO) frequency between 5~GHz and 7~GHz  to be measured. The MKIDs are designed to have resonant frequencies in range 4\ldots8~GHz with a frequency spacing that also scales with frequency. Since the central 50~MHz of the readout is not usable it takes 4 different LO tunings to measure the entire array.

To measure the position-dependent response of the arrays we use a hot source placed in the reimaged focal plane and scanned using a xy scanner. The hot source consists of a globar element placed in the focus of an enclosing elliptical mirror, shown in Fig.~\ref{fig:schema}(b). The elliptical mirror produces an image of the source at its second focus, where a $\varnothing2$~mm beam-defining aperture is placed. The output of the source is modulated at 80~Hz between 300~K and upto 1000~K by means of a rotating mirror to eliminate MKID 1/f noise and system thermal drifts. The hot source has been previously characterized to have a wide response with an effective spot size smaller than the beamsizes to be measured. To maintain a constant background on the MKIDs and to eliminate reflections, a blackened sheet significantly larger than the reimaged chip size is mounted around the source aperture. The response of the MKID as a function of the source position is measured using a step-and-integrate strategy. The typical step size of the source is chosen to be close to spatial Nyquist sampling (2.5~mm) to enable an efficient sampling of the entire field of view. For each xy point, typically one second of data is recorded. The hot source temperature is adjusted so the maximum signal during the measurement matches the MKID instantaneous dynamic range. This is  taken in MKID phase readout~\cite{day03,gao:APL07} as $\sim$1~rad. with respect to the MKID resonance circle in the complex plane. In post processing, the data is calibrated to an effective frequency shift via the MKID phase signal using the strategy outlined in~\cite{bisigello:JLTP16}. This linearizes the signal with respect to the optical power and removes responsivity changes due to drifts in the optical loading. This results in reproducible beams for different source powers, MKID readout tone frequencies and powers. The position-dependent response is determined by applying a flat-top windowed FFT to each second of MKID frequency response and taking the FFT amplitude at the chopper frequency. Since the chopper is not locked it has a slight frequency drift. To correct for this, we modulate a few off-KID readout tones using analogue electronics with the exact chopper frequency. This gives the exact chopper frequency in the measured data, enabling drift correction in post-processing. This was implemented only for the array with absorber. For the array without absorber we use the sum of all MKIDs for this purpose: the pedestal response ensures that a signal is present independent of the source position. 

The position dependent responses is shown in Fig.~\ref{fig:hot_beams}(a) for the array without absorbing mesh and in Fig.~\ref{fig:hot_beams}(b) for the array with absorbing mesh. In both figures we give the corrected response $P_{c}$, given by $P_{c}=\sqrt{P^{2}-P_n^2}$, where $ P_n^2$ is the mean value of the signal with the hot source outside of the field of view of the cryostat optics. The square is needed because the KID noise, readout noise and photon noise contributions all add in $P^{2}$, with $P$ the measured signal. We observe a significant reduction (~10~dB) in the pedestal response for the array with mesh absorber. To make this even clearer we show in Figs.~\ref{fig:radial_mean}(a)~\&~(b) the radial mean of the beam pattern. The radial mean averages the beam pattern over a circle centered on the pixel position, so improving on the signal to noise of the beam pattern. The solid (blue) line, representing the data without a mesh absorber, is 10~dB above the (red) dashed line, representing the data with absorber, for radii in excess of 20~mm.

The measured beam pattern, as shown in Fig.~\ref{fig:hot_beams}(b), has a noise floor at $\sim-35$~dB. This is not sufficient to measure the residual response for the array with absorber. For these reasons we have performed an additional measurement using a harmonic source and a multiplier chain in a heterodyne configuration giving the phase and amplitude (P\&A) beam patterns~\cite{thomas:IEEETTST12}, similar to the method presented in Davis et al.~\cite{Davis:IEEETTT16}. The multiplier chain is used as a stationary local oscillator that is coupled to the entire the array with a thin mylar beam splitter. The harmonic mixer is scanned in the reimaged focal plane. The sources are operated at a small, $\sim 424$~Hz, offset modulating the signal. The magnitude and phase of this modulation gives the amplitude and phase of the system beam pattern, with only the amplitude presented here. The scanned positions in xy plane are chosen to be every 1~mm, less than FWHM/4, the full width half maximum beamwidth, to resolve the beam shape in more detail. Due to the dual-source modulation, the P\&A measurement measures the amplitude of the beam pattern whereas the hot source gives the power~\cite{thomas:IEEETTST12}. This means the P\&A measurement dynamic range is square of that with the hot source for the same modulation level, giving a noise floor of $\lesssim-60$~dB as shown in Fig~\ref{fig:PA_beam}. Additionally the harmonic mixer uses a thin-walled open waveguide to launch the radiation; so it is a single mode, single polarization and a single frequency point source with an near isotropic beam pattern. The pattern measured with this source can therefore directly be compared to simulations. With this extra dynamic range we clearly observe in the 2D beam pattern that even with an on-chip absorber, a residual large area response is still visible at a $-40$ to $-50$~dB level.

\section{Comparison to optical simulation}
To further assess the measured performance and to compare it to the theoretical performance, we take a POP simulation of the camera cryostat optics and compare this to the radial mean of the measured response. This is shown in Fig.~\ref{fig:radial_mean}. The POP model approximates the optics, using the 3D designs of the mirrors and, in particular, taking as the detector beam a Gaussian beam with a beamwaist of 0.72~mm. This is the radius at the $1/e^{2}$ value of the beam pattern fitted from simulations of the lens-antenna system. Note that the measured and simulated far field beam patterns of an similar lens-antenna coupled MKID are in excellent agreement \cite{Ferrari:inprep}. For the array with an absorber the P\&A measurement is in good agreement to the POP simulation down to the $\sim-35$~dB level. Further away from the beam center the measured signal is higher and we observe a clear low-level extended feature on the beam pattern. This is the case also for the array with an absorbing mesh, however here it even extends beyond the chip area as shown in Fig.~\ref{fig:PA_beam}. This deviation we refer to as a residual error beam. Given the fact that part of the residual error beam extends beyond the chip edge we can imply that part of this signal is not from the surface wave in the chip, but from contributions from the measurement set-up such as: reflections in the optical path, for example from the optical window and filters; additional Ruze~\cite{ruze66} scattering from optical surface roughness and errors; residual diffraction, not in the simulation due to simulation sampling errors or off components near the beam. Such effects will give contributions to the beam pattern over the optical field of view, matching what is observed for the array with absorber as seen in Fig.~\ref{fig:PA_beam}. It is important to note that the accuracy of the POP simulation at such low levels off-axis may be degraded due to simulation sampling and the POP algorithm itself. 

To illustrate the difference between arrays we show in Fig.~\ref{fig:radial_mean}(c) the encircled energy, which is the integral of the beam pattern over a circle centered on the pixel position. Note, the encircled energy is normalized to the simulation (POP) and corrected for small variations in the FWHM between the curves. The FWHM and hence its integral vary between measurements on the order of $\sim 10~\%$ due to the finite source size for hot source measurements, defocus, slight optical misalignments, and slight difference between measurements and simulation. Noting that the main beam response dominates up to a radius of 10~mm, the power inside this radius is used to normalize the encircled energy to the POP beam pattern. The encircled energy shows that without absorber there is almost 1.8~$\times$ the response of the array with an absorber, and that this extra power is distributed away from the pixel center. 

\section{Extended source response}
To cross check the beam pattern measurements we measured the response of a few single pixels as a function of load size for the array with an on-chip absorber. To test this the source (in Fig.~\ref{fig:schema}(a)) is replaced with a variety of 300~K sources sizes, while the rest of the field of view is allowed to pass onto a large 77~K liquid nitrogen load. Two source types are presented: large sources from Eccosorb AN absorber~\cite{eccosorb}, with sizes corresponding to the full field of view and smaller shown in Table~\ref{table:sig_size}; sub-beam size sources using metal balls mounted to 12~$\mu$m mylar strip of width $\sim$20~mm. The mylar strip is almost fully transparent, giving a small, $\sim5$~K constant background that can be clearly distinguished from the more localised metal ball signal. The metal balls block the nitrogen load and reflect into the beam 300~K from the room environment, with the effective load temperature seen by the detector dependent on the beam filling fraction. For large sources the signal from the loads was larger than the designed dynamic range ($\sim$50~K), so as a crosscheck the frequency sweep of the MKID was used to determine the actual shift of the resonant frequency versus different loads. The small sources are linearized to an effective frequency from the f-sweep as in~\cite{bisigello:JLTP16}.

\begin{table}
\center
\caption{Normalized optical response versus load size, compared to integration of beam pattern.}\label{table:sig_size}
\begin{tabular}{l l l}
Load & Measured & Integral of  \\
 &   & beam pattern\\
\hline \\[-0.9em]
300~K& 1 & 1 \\
30 mm strip  & 0.92 & 0.93\\
25x25 mm  & 0.89 & 0.85 \\
\hline
\end{tabular}
\end{table}

\begin{figure}
\centering
\includegraphics*[width=8.5cm]{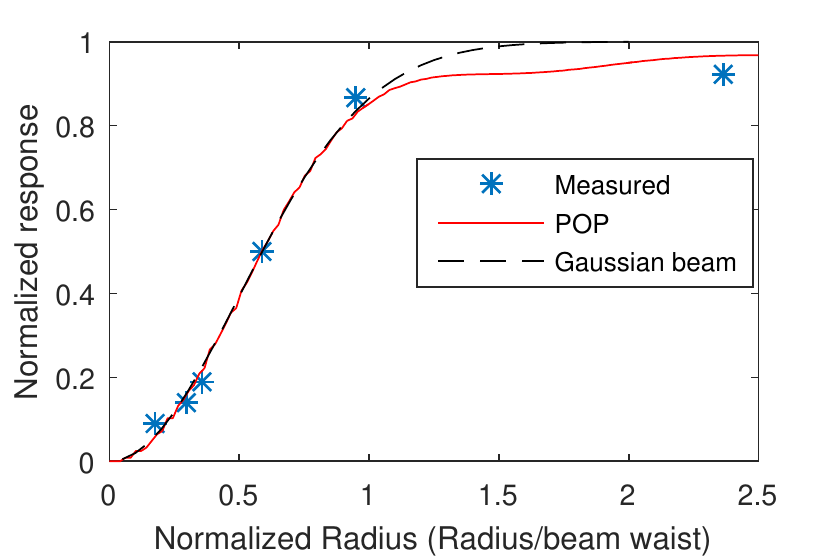}
\caption{Normalized peak response versus source size, compared to encircled energy of simulation, POP, and Gaussian beam equations~\cite{Goldsmith}.}\label{fig:response_r}
\end{figure}

For small sources, the result is shown Fig.~\ref{fig:response_r} showing a close match to integrals of the POP simulation and the Gaussian beam equations~\cite{Goldsmith}. For large sources the result is summarized in table~\ref{table:sig_size}. The response is normalized to 1 on 300~K and 0 on 77~K. We observe that there is good agreement between the response and the integration of P\&A beam pattern. These results show the beam pattern to be a complete description of the detector response, from point sources to extended sources. 

\section{Conclusions}
In this paper we have shown that a large, monolithic array of lens-antenna coupled MKIDs can respond to radiation, on a $-30$~dB level, over the entire chip area. This ‘pedestal’ response is associated with radiation scattered inside the dielectric of the array substrate, it is commonly referred to as a surface wave. The integrated response of the pedestal approaches the main beam response. Such a response destroys the imaging properties of the array, particularly for extended sources. We have shown that the surface wave can be suppressed effectively by including a matched absorbing layer in between the detector chip and the lens array, leading to a beam pattern response close to the expected spatial response from a POP model down to the $-35$~dB level. The absorbing layer reduces the surface wave by at least 10~dB. A remaining extended beam pattern feature has response outside the chip area so is therefore at least partly due to imperfections in the setup, and not associated with the detector assembly. The measured array now meets the requirements of sensitivity and beam pattern for both point and extended sources as is needed for future far infra-red, submillimeter wave (THz) and mm-wave astronomy, such as reviewed in~\cite{baselmans:AA17}.   

While lens-antenna coupled MKIDs are presented, the problem of surface waves is common for any detector system on a monolithic transparent substrate. This is a low-level effect that will become more important for large focal plane arrays or any large array requiring low pixel-pixel on chip optical crosstalk. For example, spectrometer on chip applications~\cite{endo:JLTP12,superspecLTD16short,cataldo:AO14} also require high on chip rejection of out of band and stray radiation, so similar solutions to absorb surface wave contributions are required~\cite{barrentine:SPIE16}.
 
\section*{Acknowledgment}
\addcontentsline{toc}{section}{Acknowledgment} 
 The authors would like to thank: Akira Endo and Kotaro Kohno for the use of a multiplier chain; Ronald Hesper and the NOVA ALMA group for support with the phase and amplitude beam patterns; Martin Eggens, Geert Keizer, Bert Kramer, Martin Grim, Heino Smit, Duc Nguyen, Henk Od\'e, Rob van der Schuur and Jarno Panman for technical support.

\appendix[Optical-thermal design issues]
\subsection{Chip thermalization}
The chip itself can suffer from thermalization issues due to the incident absorbed optical and readout power. At the used optical loading thermal effects are measurable, but small at 260~mK. Without a gold layer, there is a small parasitic thermal response due to changes in loading applied to the entire array of order of $\sim$ 10~\% of the single pixel response with a time constant of 18~s. This is associated with a chip heating from 260 to 277~mK in the middle of the array, which was measured in a separate cooldown with a thermometer mounted on center an array without a thermalization layer. The addition of a thermalization layer reduced the $\Delta T$ to below that measurable. At 240~mK, used in the presented measurements, with the gold thermalization layer there is no measurable residual parasitic thermal response associated with changes in the full array optical loading. 
\subsection{Optical and infra-red filtering}\label{app:B}
\begin{table}
\centering
\caption{Filter stack overview.}\label{table:filters}
\begin{tabular}{l l l}
Name & Position and nominal T & Size  \\
\hline \\[-0.9em]
HDPE window & Window 300~K & $\varnothing$170~mm \\
Scatterer & 300~K & $140\times$150~mm \\
Shader $15\mu$m  & 50~K &$ \varnothing$200~mm \\
Scatterer & 50~K & $\varnothing$200~mm \\
Shader $15\mu$m  & 50~K &  $\varnothing$200~mm \\
LP 3~THz & 50 K & $\varnothing$200~mm \\
Shader($\times$2) $30\mu$m  & Window 4~K & $\varnothing$210~mm \\
LP 1.1~THz & Window 4~K & $\varnothing$210~mm \\
HDPE 8mm thick & Pupil 4~K & $\varnothing$160~mm \\
LP 400~GHz & 800~mK & $110\times$110~mm\\
BP 350~GHz & 240~mK & $75\times75$~mm \\
\hline
\end{tabular}
\end{table}
The optical measurement band is defined by a set of optical filters and requires $>60$~dB rejection of out of band radiation. A particular problem is that the filters are large and made of plastic (typically mylar) so are poorly thermalized. This requires additional filters to block the reradiated heat from the hot filters~\cite{Tucker:SPIE006}. The total filter stack~\cite{QMC} consists of (see table~\ref{table:filters}): infra-red (IR) scatterers, that scatter near-IR radiation and are used to reduce window condensation; IR shaders, thin film reflective near to mid-IR low-pass filters, defined by their low-pass wavelength, these reflect most of the out of band radiation power back out of the cryostat window; metal mesh low-pass filters (LP), that reflect far-IR and out of band submillimeter wave radiation, but absorb mid-IR; a band-pass (BP) filter to define the measurement band; an 8~mm thick high-density polyethylene (HDPE) sheet is used as the cryostat optical window; an additional 8mm HDPE sheet is used to absorb residual reradiated IR radiation inside the cryostat.

\bibliographystyle{IEEEtran}
\bibliography{pedestalv13arXiv.bbl}

\end{document}